\documentclass[epj]{svjour}
\usepackage{amsfonts}
\usepackage{amsmath}
\usepackage{graphicx}
\usepackage{subfigure}
\usepackage{dcolumn}
\usepackage{bm}
\usepackage{booktabs}
\usepackage{color}

\newcommand{\figcaption}{\def\@captype{figure}\caption}
\newcommand{\tabcaption}{\def\@captype{table}\caption}

\newcommand{\Rmnum}[1]{\expandafter\@slowromancap\romannumeral #1@}
\def\hlinewd#1{%
  \noalign{\ifnum0=`}\fi\hrule \@height #1 \futurelet
   \reserved@a\@xhline}
\makeatother

\def\FF(s){\left[(\alpha+\beta)m_c^2-\alpha\beta s\right]}
\def\HH(s){\left[m_c^2-\alpha(1-\alpha) s\right]}

\allowdisplaybreaks[4]

\usepackage{ulem}

\usepackage{graphics}

\begin{document}

\title{Understanding the internal structures of
the $X(4140)$, $X(4274)$, $X(4500)$ and $X(4700)$}

\author{Hua-Xing Chen\inst{1}
\and Er-Liang Cui\inst{1}
\and Wei Chen\inst{2}
\thanks{wec053@mail.usask.ca}
\and Xiang Liu\inst{3,4}
\thanks{xiangliu@lzu.edu.cn}
\and Shi-Lin Zhu\inst{5,6,7}
\thanks{zhusl@pku.edu.cn}
}                     
\offprints{}          
\institute{
School of Physics and Beijing Key Laboratory of Advanced Nuclear Materials and Physics, Beihang University, Beijing 100191, China
\and
Department of Physics and Engineering Physics, University of Saskatchewan, Saskatoon, Saskatchewan, S7N 5E2, Canada
\and
School of Physical Science and Technology, Lanzhou University, Lanzhou 730000, China
\and
Research Center for Hadron and CSR Physics, Lanzhou University and Institute of Modern Physics of CAS, Lanzhou 730000, China
\and
School of Physics and State Key Laboratory of Nuclear Physics and Technology, Peking University, Beijing 100871, China
\and
Collaborative Innovation Center of Quantum Matter, Beijing 100871, China
\and
Center of High Energy Physics, Peking University, Beijing 100871, China}
\date{Received: date / Revised version: date}
%
\abstract{
We investigate the newly observed $X(4500)$ and $X(4700)$ based on the diquark-antidiquark configuration within the framework of QCD sum rules. Both of them may be interpreted as the $D$-wave $cs\bar{c}\bar{s}$ tetraquark states of $J^P = 0^+$, but with opposite color structures, which is remarkably similar to the result obtained in Ref.~\cite{Chen:2010ze} that the $X(4140)$ and $X(4274)$ can be both interpreted as the $S$-wave $cs\bar{c}\bar{s}$ tetraquark states of $J^P = 1^+$, also with opposite color structures. However, the extracted masses and these suggested assignments to these $X$ states do depend on these running quark masses where $m_s (2 \mbox{ GeV}) = 95 \pm 5$ MeV and $m_c (m_c) = 1.23 \pm 0.09$ GeV. As a byproduct, the masses of the hidden-bottom partner states of the $X(4500)$ and $X(4700)$ are extracted to be both around 10.64 GeV, which can be searched for in the $\Upsilon \phi$ invariant mass distribution.
} 
\maketitle

{\it Introduction.}---It is well known that our world is made from
nucleons and electrons while nucleons are made from quarks and
gluons. However, we still know little (not enough) on how quarks and
gluons compose nucleons, which can be better understood by exploring
exotic matter beyond the conventional quark model, such as
glueballs, hybrids and multiquark states,
etc.~\cite{Agashe:2014kda,Jaffe:2004ph,Liu:2013waa,Chen:2016qju}.
With significant experimental progress over the past decade, lots of
multiquark candidates have been observed, including dozens of
charmonium-like and bottomonium-like $XYZ$ states~\cite{Agashe:2014kda} and
the hidden-charm pentaquark states $P_c(4380)$ and
$P_c(4450)$~\cite{Aaij:2015tga}. They are new blocks of QCD matter,
and provide important hints to deepen our understanding of the
non-perturbative quantum chromodynamics (QCD).

Very recently the LHCb Collaboration confirmed the $X(4140)$ and
$X(4274)$ in the $J/\psi \phi$ invariant mass distribution and
determined their spin-parity quantum numbers to be both $J^P =
1^{++}$~\cite{lhcb}. At the same time they investigated the high $J/\psi \phi$ mass region
for the first time, where the results can be described as a nonresonant term plus
two new $J^P = 0^{++}$ resonances, named as the $X(4500)$ and $X(4700)$. Their masses and widths
were measured to be:
\begin{flalign}
\nonumber & X(4140) : M = 4146.5 \pm 4.5 ^{+4.6}_{-2.8} \mbox{ MeV}
\, ,
\\ \nonumber & ~~~~~~~~~~~~~~ \Gamma = 83 \pm 21 ^{+21}_{-14} \mbox{ MeV} \, ,
\\
\nonumber & X(4274) : M = 4273.3 \pm 8.3 ^{+17.2}_{-3.6} \mbox{ MeV}
\, ,
\\ \nonumber & ~~~~~~~~~~~~~~ \Gamma = 56 \pm 11 ^{+8}_{-11} \mbox{ MeV} \, ,
\\
\nonumber & X(4500) : M = 4506 \pm 11 ^{+12}_{-15} \mbox{ MeV} \, ,
\\ \nonumber & ~~~~~~~~~~~~~~ \Gamma = 92 \pm 21 ^{+21}_{-20} \mbox{ MeV} \, ,
\\
\nonumber & X(4700) : M = 4704 \pm 10 ^{+14}_{-24} \mbox{ MeV} \, ,
\\ \nonumber & ~~~~~~~~~~~~~~ \Gamma = 120 \pm 31 ^{+42}_{-33} \mbox{ MeV} \, .
\end{flalign}
The $X(4140)$~\cite{Aaltonen:2009tz} and the
$X(4274)$~\cite{Aaltonen:2011at} were first reported by the CDF
Collaboration in 2009 and 2011 respectively. Many theoretical
explanations were proposed such as the $D_s^* \bar D_s^*$ and $D_s
\bar D_{s0}(2317)$ molecular
states~\cite{Liu:2008tn,Liu:2009ei,Mahajan:2009pj,Branz:2009yt,Ding:2009vd,Liu:2010hf,Wang:2011uk,Finazzo:2011he,He:2011ed,HidalgoDuque:2012pq},
compact tetraquark states (diquark-antidiquark
states)~\cite{Stancu:2009ka,Patel:2014vua}, dynamically generated
resonances~\cite{Molina:2009ct,Branz:2010rj}, and coupled-channel
effects~\cite{Danilkin:2009hr,vanBeveren:2009dc}, etc.

Among these studies, the results obtained within the framework of
QCD sum rules are significant
~\cite{Chen:2010ze,Albuquerque:2009ak,Zhang:2009st,Wang:2009ue,Wang:2009ry}, which method has
been applied to studied many other multiquark candidates~\cite{Chen:2006zh,Chen:2015moa,Chen:2016ymy}.
In 2010, Chen, {\it et al.} studied the vector and axial-vector
charmonium-like states systematically in Ref. ~\cite{Chen:2010ze},
where they used the following two $J^P = 1^+$ currents to perform
QCD sum rule analyses ($a$ and $b$ are color indices):
\begin{eqnarray}
J_{3\mu} &=& s^T_a C\gamma_5 c_b (\bar{s}_a \gamma_{\mu}C
\bar{c}^T_b + \bar{s}_b \gamma_{\mu}C \bar{c}^T_a) \label{def:j3}
\\ \nonumber && ~~~~~~ ~~~~~~ + s^T_a C\gamma_{\mu} c_b (\bar{s}_a \gamma_5C \bar{c}^T_b + \bar{s}_b \gamma_5C \bar{c}^T_a) \, ,
\\
J_{4\mu} &=& s^T_a C\gamma_5 c_b (\bar{s}_a \gamma_{\mu}C
\bar{c}^T_b - \bar{s}_b \gamma_{\mu}C \bar{c}^T_a) \label{def:j4}
\\ \nonumber && ~~~~~~ ~~~~~~ + s^T_a C\gamma_{\mu} c_b (\bar{s}_a \gamma_5C \bar{c}^T_b - \bar{s}_b \gamma_5C \bar{c}^T_a) \, ,
\end{eqnarray}
which are constructed using diquark and antidiquark fields. There
are altogether five diquark fields: $q^T_a C q_b$, $q^T_a C\gamma_5
q_b$, $q^T_a C\gamma_\mu q_b$, $q^T_a C\gamma_\mu\gamma_5 q_b$ and
$q^T_a C\sigma_{\mu\nu} q_b$~\cite{Chen:2006hy,Chen:2007xr}. Among
them, the $S$-wave diquark fields $q_a^T C\gamma_5 q_b$ and $q_a^T
C\gamma_{\mu} q_b$ are
favored~\cite{Jaffe:2004ph,Maiani:2004vq,Kleiv:2013dta}, which can
be used to further construct the ``good'' and ``bad'' diquarks by
demanding their color structure to be antisymmetric $[\mathbf{\bar
3_c}]_{qq}$ (by simply adding a totally antisymmetric tensor
$\epsilon^{abc}$)~\cite{Jaffe:2004ph}. The other three ``worse''
diquarks all contain $P$-wave components~\cite{Jaffe:2004ph}.

The current $J_{4\mu}$ defined in Eq.~(\ref{def:j4}) has the
antisymmetric color structure $[\mathbf{\bar 3_c}]_{cs} \otimes
[\mathbf{3_c}]_{\bar{c}\bar s}$. Hence, this interpolating current
consists of one ``good'' diquark and one ``bad'' antidiquark, and is
the most favored one among all the $J^P = 1^+$ currents. Its
extracted mass is $4.07 \pm 0.10$ GeV~\cite{Chen:2010ze}, consistent
with the experimental mass of the $X(4140)$~\cite{Agashe:2014kda}.
The current $J_{3\mu}$ defined in Eq.~(\ref{def:j3}) consists of one
similar diquark and one similar antidiquark, but having the
symmetric color structure $[\mathbf{6_c}]_{cs} \otimes [\mathbf{
\bar 6_c}]_{\bar{c}\bar s}$. Hence, it is less favored but still
better than other currents containing ``worse
''diquarks~\cite{Jaffe:2004ph}. Its extracted mass is $4.22 \pm
0.10$ GeV~\cite{Chen:2010ze}, consistent with the experimental mass
of the $X(4274)$~\cite{Agashe:2014kda}.

The P-wave vector tetraquark states were discussed extensively in Ref. \cite{Lebed:2016yvr}.
There also exist investigations of the scalar tetraquark states. In
Refs.~\cite{Albuquerque:2009ak,Zhang:2009st,Wang:2009ue,Wang:2009ry}
three groups studied the scalar $D^*_s \bar D^*_s$ molecular state
through the current composed of two vector meson fields
\begin{eqnarray}
J_{D^*_s \bar D^*_s}(x) = \bar c_a(x) \gamma_\mu s_a(x) \bar s_b(x)
\gamma^\mu c_b(x) \, . \label{def:jDss}
\end{eqnarray}
Two of the three groups obtained similar results using $J_{D^*_s
\bar D^*_s}$, $4.14 \pm 0.09$ GeV ~\cite{Albuquerque:2009ak} and
$4.13 \pm 0.10$ GeV ~\cite{Zhang:2009st}. The extracted mass is
$3.91 \pm 0.10$ GeV with the $D_s \bar D_s$ current
~\cite{Zhang:2009st}
\begin{eqnarray}
J_{D_s \bar D_s}(x) = \bar c_a(x) \gamma_5 s_a(x) \bar s_b(x)
\gamma_5 c_b(x) \, . \label{def:jDs}
\end{eqnarray}
The third group extracted a significantly larger mass $4.43 \pm
0.16$ GeV~\cite{Wang:2009ue,Wang:2009ry}, which is significantly
larger than the $D_s^* \bar D_s^*$ threshold, 4.22 GeV.

The $X(4500)$ and $X(4700)$ have masses significantly larger than
the $X(4140)$ and $X(4274)$ of $J^P = 1^+$. They can be good
candidates of the $D$-wave tetraquark states of $J^P = 0^+$, whose
possible angular momenta are $\{[c s]_{s=1}$ $[\bar c \bar
s]_{s=1}; L = S = 2, J = 0\}$. Hence, the ``good'' diquark of
$S=0$ can not be used, but they can still be composed by the ``bad''
diquark of $S=1$, $\epsilon^{abc} q_a^T C\gamma_{\mu} q_b$. We also
need the $P$-wave ``bad'' diquark field $\epsilon^{abc} q^T_a C
\gamma_{\mu_1} D_{\mu_2} q_b$ and the $D$-wave ``bad'' diquark field
$\epsilon^{abc} q^T_a C \gamma_{\mu_1} D_{\mu_2}D_{\mu_3} q_b$, as
well as their partners having the symmetric color structure
$[\mathbf{6_c}]_{qq}$.

In this work we will show that the $X(4500)$ and $X(4700)$ can be
both interpreted as $D$-wave tetraquark states with the quark
content $cs\bar{c}\bar{s}$ and $J^P = 0^+$: the $X(4500)$ consists
of one $D$-wave ``bad'' diquark and one $S$-wave ``bad''
antidiquark, having the antisymmetric color structure $[\mathbf{\bar
3_c}]_{cs} \otimes [\mathbf{3_c}]_{\bar{c}\bar s}$;
the $X(4700)$ consists of one similar $D$-wave diquark and one
similar $S$-wave antidiquark, but having the symmetric color
structure $[\mathbf{6_c}]_{cs} \otimes [\mathbf{ \bar
6_c}]_{\bar{c}\bar s}$.

These two interpretations are remarkably similar to those obtained
in Ref.~\cite{Chen:2010ze} that the $X(4140)$ and $X(4274)$ can be
both interpreted as $S$-wave tetraquark states with the quark
content $cs\bar{c}\bar{s}$ and $J^P = 0^+$: the $X(4140)$ consists
of one $S$-wave ``good'' diquark and one $S$-wave ``bad''
antidiquark, having the antisymmetric color structure $[\mathbf{\bar
3_c}]_{cs} \otimes [\mathbf{3_c}]_{\bar{c}\bar s}$; the $X(4274)$
consists of two similar $S$-wave diquarks, but having the symmetric
color structure $[\mathbf{6_c}]_{cs} \otimes [\mathbf{ \bar
6_c}]_{\bar{c}\bar s}$.

To examine these interpretations, we investigate the bottom partner
states of the $X(4500)$ and $X(4700)$, and extract their masses to
be both around 10.64 GeV. We propose to search for them in the
$\Upsilon \phi$ invariant mass distribution with the running of LHC
at 13 TeV and forthcoming BelleII. If the above interpretations of
the $X(4140)$, $X(4274)$, $X(4500)$ and $X(4700)$ are correct, their
dual partner would be quite interesting, such as the $S$-wave scalar
and the $D$-wave axial-vector $cs\bar{c}\bar{s}$ tetraquark states,
consisting of two ``bad'' diquarks with both symmetric and
antisymmetric color structures. All their related studies, both
experimentally and theoretically, can deepen our understanding of
the non-perturbative QCD. Especially, our present study can be
helpful to improve our understanding of the internal structures of
exotic hadrons.

{\it Interpretation of the $X(4500)$ and $X(4700)$.}---As the first
step, we use the $S/P/D$-waves ``axial-vector'' diquarks of $S=1$ to
construct the $D$-wave $cs\bar{c}\bar{s}$ tetraquark currents of
$J^P = 0^+$. There are two possible ways. One way is to use the
combination of one $P$-wave diquark and one $P$-wave antidiquark:
\begin{eqnarray}
\nonumber J_{1\pm} &=& c_a^T C \gamma_{\mu_1} [D_{\mu_3} s_b]
(\bar{c}_a \gamma_{\mu_2} C [D^\dag_{\mu_4} \bar{s}_b^T] \pm
\bar{c}_b \gamma_{\mu_2} C [D^\dag_{\mu_4} \bar{s}_a^T])
\\ && \times \left( g^{\mu_1 \mu_3} g^{\mu_2 \mu_4} + g^{\mu_1 \mu_4} g^{\mu_2 \mu_3} - g^{\mu_1 \mu_2} g^{\mu_3 \mu_4}/2 \right) \, ,
\label{def:j1pm}
\end{eqnarray}
where $J_{1+}$ has the symmetric color structure
$[\mathbf{6_c}]_{cs} \otimes [\mathbf{ \bar 6_c}]_{\bar{c}\bar s}$,
and $J_{1-}$ has the antisymmetric color structure $[\mathbf{\bar
3_c}]_{cs} \otimes [\mathbf{3_c}]_{\bar{c}\bar s}$; they both have
$l_{cs} = l_{\bar c \bar s} = 1$ and $s_{cs} = s_{\bar c \bar s} =
1$, and their total momenta are $L=S=2$ and $J=0$.

The other way is to use the combination of one $D$-wave diquark and
one $S$-wave antidiquark:
\begin{eqnarray}
\nonumber J_{2\pm} &=& c_a^T C \gamma_{\mu_1} [D_{\mu_3}D_{\mu_4}
s_b] (\bar{c}_a \gamma_{\mu_2} C \bar{s}_b^T \pm \bar{c}_b
\gamma_{\mu_2} C \bar{s}_a^T)
\\ && \times \left( g^{\mu_1 \mu_3} g^{\mu_2 \mu_4} + g^{\mu_1 \mu_4} g^{\mu_2 \mu_3} - g^{\mu_1 \mu_2} g^{\mu_3 \mu_4}/2 \right) \, ,
\label{def:j2pm}
\end{eqnarray}
where $J_{2+}$ has the symmetric color structure
$[\mathbf{6_c}]_{cs} \otimes [\mathbf{ \bar 6_c}]_{\bar{c}\bar s}$,
and $J_{2-}$ has the antisymmetric color structure $[\mathbf{\bar
3_c}]_{cs} \otimes [\mathbf{3_c}]_{\bar{c}\bar s}$; they both have
$l_{cs} = 2$, $l_{\bar c \bar s} = 0$ and $s_{cs} = s_{\bar c \bar
s} = 1$, and their total momenta are also $L=S=2$ and $J=0$.

The $D$-wave tetraquark currents can also be constructed by using
the $S/P/D$-waves mesonic fields,
\begin{eqnarray}
J_1^\prime &=& \bar{c}_a \gamma_{\mu_1} [D_{\mu_3} s_a]
[D^\dag_{\mu_4} \bar{s}_b] \gamma_{\mu_2} c_b
\\ \nonumber && ~~~ \times \left( g^{\mu_1 \mu_3} g^{\mu_2 \mu_4} + g^{\mu_1 \mu_4} g^{\mu_2 \mu_3} - g^{\mu_1 \mu_2} g^{\mu_3 \mu_4}/2 \right) \, ,
\\ J_2^\prime &=& \bar{c}_a \gamma_{\mu_1} [D_{\mu_3}D_{\mu_4} s_a] \bar{s}_b \gamma_{\mu_2} c_b
\\ \nonumber && ~~~ \times \left( g^{\mu_1 \mu_3} g^{\mu_2 \mu_4} + g^{\mu_1 \mu_4} g^{\mu_2 \mu_3} - g^{\mu_1 \mu_2} g^{\mu_3 \mu_4}/2 \right) \, .
\end{eqnarray}
These currents have the color structure $[\mathbf{1_c}]_{\bar cs} \otimes [\mathbf{1_c}]_{\bar{s} c}$.
Moreover, the color-octet quark-antiquark pairs can also be used to construct
the tetraquark currents having the hidden-color structure $[\mathbf{8_c}]_{\bar cs} \otimes [\mathbf{8_c}]_{\bar{s} c}$.
We will not investigate such currents in the present study, but note that we can use
the Fierz and color rearrangements to relate the local diquark-antiquark and dimeson currents (see Refs.~\cite{Chen:2016qju,Chen:2006hy,Chen:2007xr} for detailed discussions).

In the following we use the currents $J_{i\pm}$ ($i=1,2$) to study
the $D$-wave $cs\bar{c}\bar{s}$ tetraquark states of $J^P = 0^+$,
denoted as $X$, using the method of QCD sum rules, which provides a
model-independent method to study nonperturbative problems in strong
interaction
physics~\cite{Shifman:1978bx,Reinders:1984sr,Nielsen:2009uh,colangelo,Narison:2002pw}.
We need to deal with the two derivative operators inside $J_{i\pm}$,
which has been applied to study the $D$ and $F$-wave heavy-light
mesons~\cite{Zhou:2014ytp,Zhou:2015ywa,Chen:2011qu}.
$J_{i\pm}$ couples to $X$ through
\begin{eqnarray}
\langle0| J |X\rangle = f_X \, . \label{coupling parameter}
\end{eqnarray}
Then the two-point correlation function can be written as
\begin{align}
\Pi(p^{2})&= i\int d^4x e^{i p \cdot
x}\langle0|T[J(x)J^{\dag}(0)]|0\rangle \, , \label{equ:Pi}
\end{align}
which can be calculated in the QCD operator product expansion (OPE)
up to certain order in the expansion, and then matched with a
hadronic parametrization to extract information about $X$.

At the hadron level, Eq.~(\ref{equ:Pi}) can be written as
%
\begin{equation}
\Pi(p^2)={\frac{1}{\pi}}\int^\infty_{s_<}\frac{{\rm Im}
\Pi(s)}{s-p^2-i\varepsilon}ds \, , \label{eq:disper}
\end{equation}
%
where $s_<$ is the physical threshold. We define its imaginary part
as the spectral function $\rho(s)$, and evaluate it by inserting
intermediate hadron states $\sum_n|n\rangle\langle n|$
%
\begin{eqnarray}
\nonumber \rho(s) \equiv \frac{1}{\pi}{\rm Im}\Pi(s) &=&
\sum_n\delta(s-M^2_n)\langle 0|\eta|n\rangle\langle
n|{\eta^\dagger}|0\rangle
\\ &=& f_X^2\delta(s-m_X^2)+ \mbox{continuum}\, ,
\label{eq:rho}
\end{eqnarray}
%
where we only take into account the lowest-lying resonance
$|X\rangle$, and $m_X$ and $f_X$ are its mass and coupling constant,
respectively.

We can also evaluate the spectral density $\rho(s)$ at the quark and
gluon level via the QCD operator product expansion. In this work we
evaluate it up to dimension ten, including the perturbative term,
the quark condensate $\langle \bar s s \rangle$, the gluon
condensate $\langle g_s^2 GG \rangle$, the quark-gluon mixed
condensates $\langle g_s \bar s \sigma G s \rangle$ and $\langle g_s
\bar s \sigma G s \rangle^2$. The full expressions are lengthy and
will not be shown here. We have also calculated the condensates
$\langle \bar s s \rangle^2$ and $\langle \bar s s \rangle\langle
g_s \bar s \sigma G s \rangle$, which can be important in sum rule
studies~\cite{Chen:2010ze}. However, both of them vanish when the
currents $J_{i\pm}$ ($i=1\cdots2$) are used.

After performing the Borel transform at both the hadron and QCD
levels, we can express the two-point correlation function as
%
\begin{equation}
\Pi^{(all)}(M_B^2)\equiv\mathcal{B}_{M_B^2}\Pi(p^2) =
\int^\infty_{s_<} e^{-s/M_B^2} \rho(s) ds \, . \label{eq:borel}
\end{equation}
%
Then assuming the contribution from continuum states can be
approximated well by the OPE spectral density above a threshold
value $s_0$ (duality)
%
\begin{equation}
\Pi(s_0, M_B^2) = \int^{s_0}_{s_<} e^{-s/M_B^2} \rho(s) ds \, ,
\label{eq:borel}
\end{equation}
%
we finally arrive at the sum rule relation:
%
\begin{eqnarray}
M^2_X(s_0, M_B) 
&=& {\int^{s_0}_{s_<} e^{-s/M_B^2} \rho(s) s ds \over
\int^{s_0}_{s_<} e^{-s/M_B^2} \rho(s) ds} \, . \label{eq:mass}
\end{eqnarray}
%
To perform numerical analysis, we use the following QCD parameters
of quark masses and various QCD
condensates~\cite{Agashe:2014kda,colangelo,Narison:2002pw,Eidemuller:2000rc,Gimenez:2005nt,Jamin:2002ev,Ioffe:2002be,Ovchinnikov:1988gk}:
\begin{eqnarray}
\nonumber && \langle \bar q q \rangle = - (0.24 \pm 0.01)^3 \mbox{
GeV}^3 \, ,
\\ \nonumber && \langle \bar s s \rangle = (0.8\pm0.1) \times \langle \bar q q \rangle \, ,
\\ \nonumber &&\langle g_s^2GG\rangle =(0.48 \pm 0.14) \mbox{ GeV}^4\, ,
\\ \label{paramaters} && \langle g_s \bar s \sigma G s \rangle = - M_0^2 \times \langle \bar s s \rangle\, ,
\\ \nonumber && M_0^2 = 0.8 \mbox{ GeV}^2\, ,
\\ \nonumber && m_s (2 \mbox{ GeV}) = 95 \pm 5 \mbox{ MeV} \, ,
\\ \nonumber && m_c (m_c) = 1.23 \pm 0.09 \mbox{ GeV} \, ,
\end{eqnarray}
in which $\overline m_s$ and $\overline m_c$ are the ``running
masses'' of the strange and charm quarks in the $\overline{\rm MS}$
scheme. We note that there is an additional minus sign in mixed
condensates due to the different definition of the coupling constant
$g_s$ compared to that in Ref.~\cite{Reinders:1984sr}.

\begin{figure*}[hbt]
\begin{center}
\scalebox{0.6}{\includegraphics{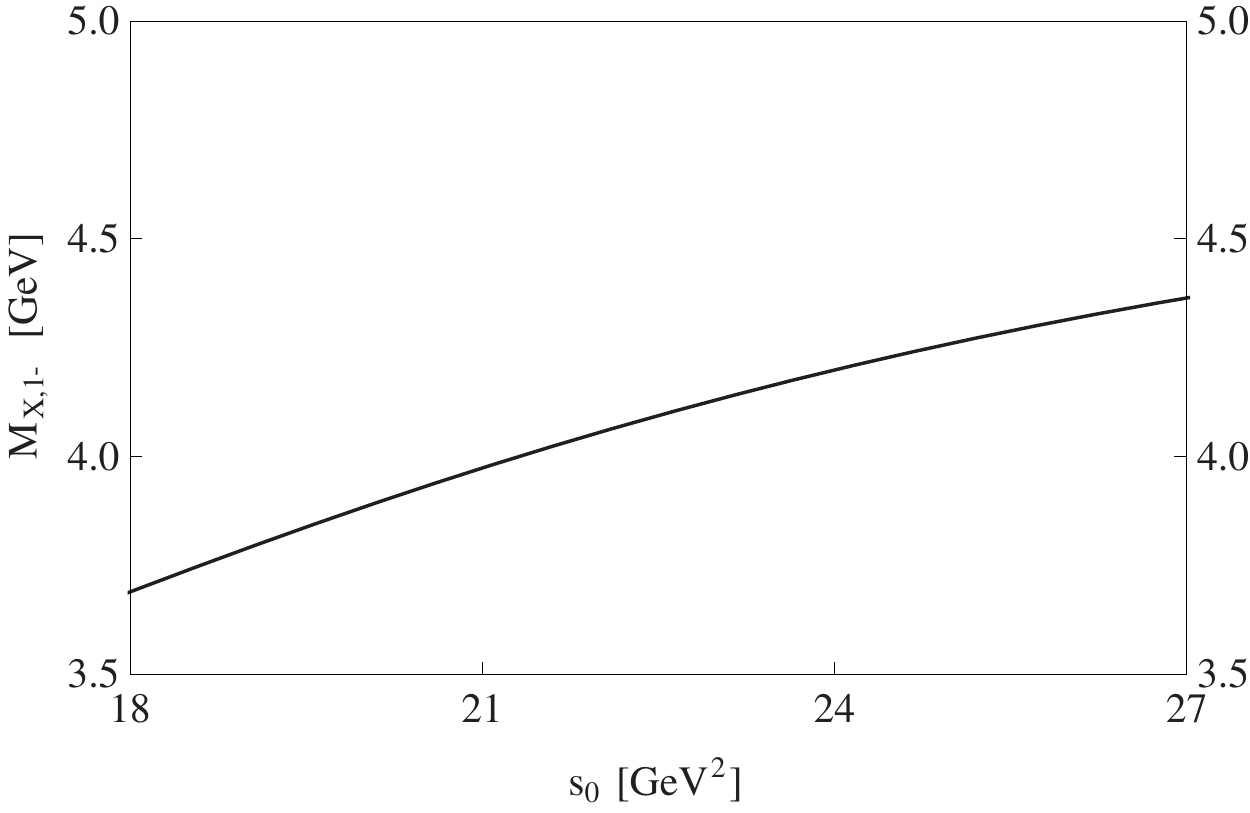}}
\scalebox{0.6}{\includegraphics{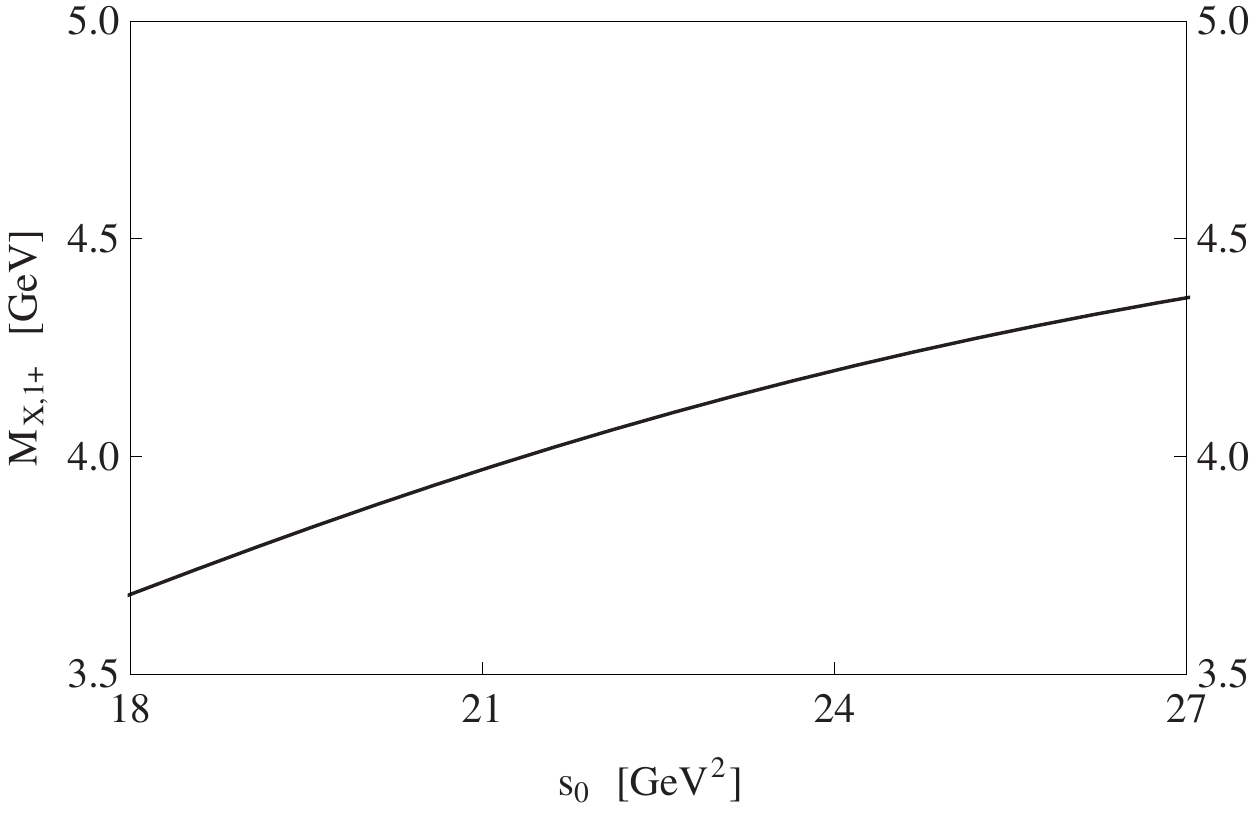}} \caption{Variations
of $M_X$ with respect to the threshold value $s_0$, when the Borel
mass $M_B$ is fixed to be $M_B^2 = 2.0$ GeV$^2$, obtained using the
currents $J_{1-}$ (left) and $J_{1+}$ (right).} \label{fig:j1}
\end{center}
\end{figure*}

There are two free parameters in Eq.~(\ref{eq:mass}): the threshold
value $s_0$ and the Borel mass $M_B$. The QCD sum rule prediction of
the hadron mass $M_X$ is only significant and reliable in suitable
regions of the parameter space $(s_0, M_B^2)$.

First we fix $M_B^2 = 2.0$ GeV$^2$ and investigate the $s_0$
dependence. The mass curves obtained using $J_{1-}$ and $J_{1+}$
(consisting of one $P$-wave diquark and one $P$-wave antidiquark)
are shown in Fig.~\ref{fig:j1}. We find that their results are
similar to each other, i.e., the evaluated masses $M_{X,1\pm}$
monotonically increase with $s_0$. We do not want conclude that this
is ``bad'' sum rule results, but it seems difficult to extract the
hadron mass $M_X$ using these two currents. Hence, we shall not
discuss $J_{1-}$ and $J_{1+}$ any more.

\begin{figure*}[hbt]
\begin{center}
\scalebox{0.6}{\includegraphics{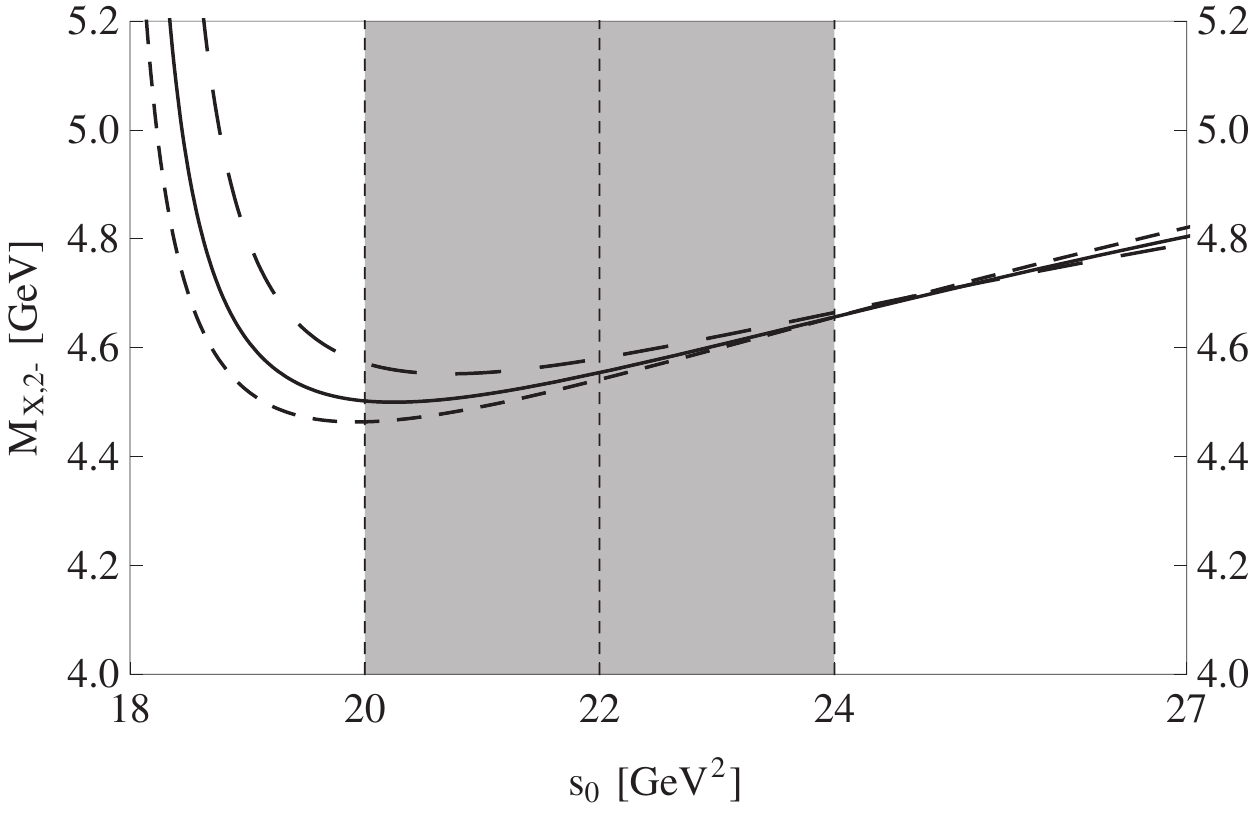}}
\scalebox{0.6}{\includegraphics{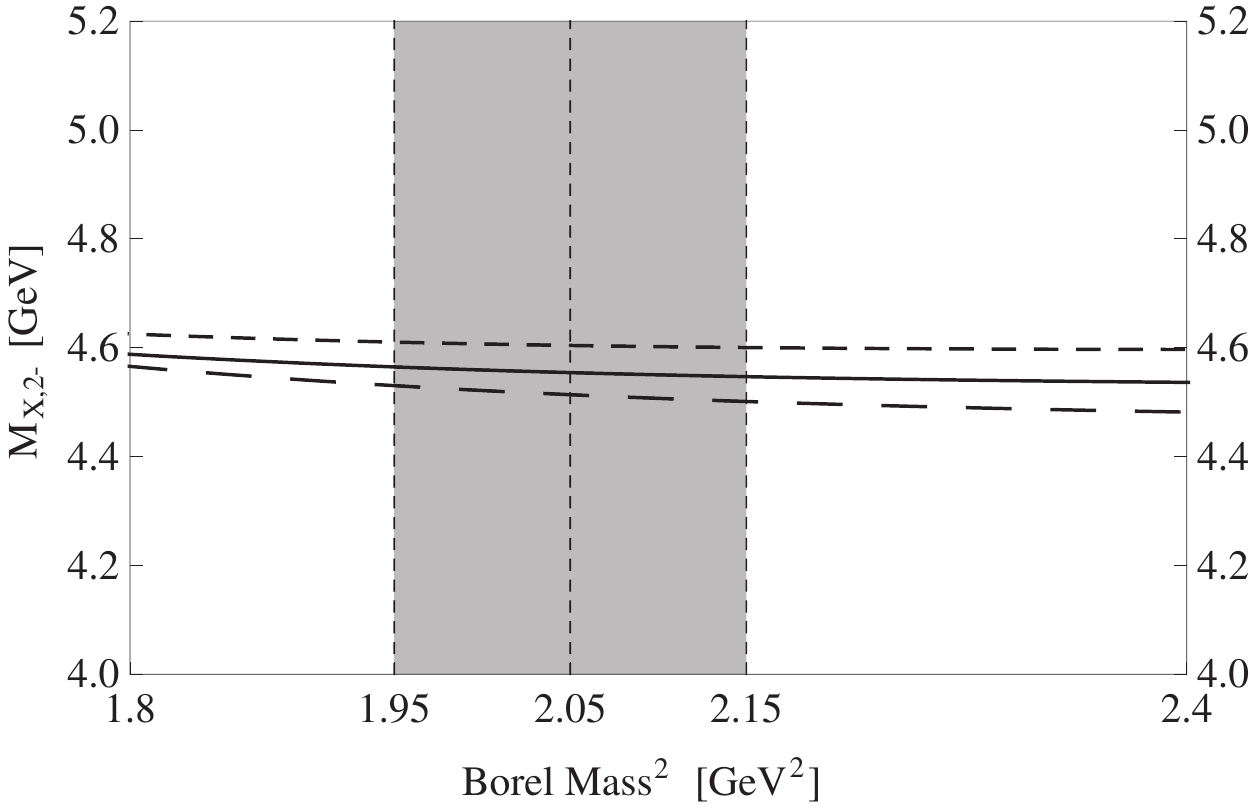}} \caption{The
variation of $M_X$ with respect to the threshold value $s_0$ (left)
and the Borel mass $M_B$ (right), calculated using the current
$J_{2-}$ of $J^P = 0^+$. In the left figure, the long-dashed, solid
and short-dashed curves are obtained by fixing $M_B^2 = 1.85$,
$2.05$ and $2.25$ GeV$^2$, respectively. In the right figure, the
long-dashed, solid and short-dashed curves are obtained for $s_0 =
21$, $22$ and $23$ GeV$^2$, respectively.} \label{fig:j2n}
\end{center}
\end{figure*}

The results obtained using $J_{2-}$ and $J_{2+}$ (consisting of one
$D$-wave diquark and one $S$-wave antidiquark) are also similar to
each other but different from $J_{1\pm}$, i.e., the obtained masses
$M_{X,2\pm}$ both have a mass plateau, where the $s_0$ dependence is
the weakest~\cite{Narison:1996fm,Chen:2010ze}. We use the current
$J_{2-}$ as an example and show the mass curves in the left panel of
Fig.~\ref{fig:j2n} as a function of the threshold value $s_0$. We
notice that the $s_0$ dependence is the weakest around $s_0 \sim 20$
GeV$^2$, and the $M_B$ dependence is the weakest around $s_0 \sim
24$ GeV$^2$. Accordingly, we choose the region $20$ GeV$^2\leq
s_0\leq 24$ GeV$^2$ as our working region, where the $s_0$ and $M_B$
dependence is both acceptable. This is our first criterion to
determine $s_0$, i.e., the $s_0$ and $M_B$ stability.

After fixing $s_0$, we use two extra criteria to constrain the Borel
mass $M_B$: a) to insure the convergence of the OPE series, we
require that the mixed condensate $\langle g_s \bar s \sigma G s
\rangle$ be less than 30\% to determine its lower limit $M_B^{min}$
(the contribution from the highest condensate $\langle g_s \bar s
\sigma G s \rangle^2$ is negligible, so we do not use it in this
criterion):
%
\begin{equation}
\label{eq_convergence} \mbox{Convergence (CVG)} \equiv \left|\frac{
\Pi_{\langle g_s \bar s \sigma G s \rangle}(\infty, M_B) }{
\Pi(\infty, M_B) }\right| \leq 30\% \, ;
\end{equation}
%
b) to insure that the one-pole parametrization in Eq.~(\ref{eq:rho})
is valid, we require that the pole contribution (PC) be larger than
20\% to determine the upper limit on $M_B^2$:
%
\begin{equation}
\label{eq_pole} \mbox{PC} \equiv \frac{ \Pi(s_0, M_B) }{ \Pi(\infty,
M_B) } \geq 20\% \, .
\end{equation}
%
The small pole contribution is due to the large powers of $s$ in the
spectral function (see other sum rule analyses for the six-quark
state $d^*(2380)$~\cite{Chen:2014vha} and the $F$-wave heavy
mesons~\cite{Zhou:2015ywa}).

Using these two criteria we obtain the working region of the Borel
mass to be $1.95$ GeV$^2< M_B^2 < 2.15$ GeV$^2$ for the current
$J_{2-}$ with $s_0 = 22$ GeV$^2$ (there exist Borel windows only
when $s_0 \geq 22$ GeV$^2$).
The variation of $M_X$ with respect to the Borel mass $M_B$ is shown
in the right panel of Fig.~\ref{fig:j2n}, where the mass curves are
very stable not only inside this Borel window but also in a larger
nearby area.

Together we obtain the working regions for the current $J_{2-}$ to
be $20$ GeV$^2\leq s_0\leq 24$ GeV$^2$ and $1.95$ GeV$^2< M_B^2 <
2.15$ GeV$^2$, where $M_X$ can be extracted to be:
\begin{eqnarray}
M_{X,2-} = 4.55^{+0.19}_{-0.13} \mbox{ GeV} \, , \label{mass2-}
\end{eqnarray}
Here the central value corresponds to $M_B^2 = 2.05$ GeV$^2$ and
$s_0 = 22$ GeV$^2$, and the uncertainty comes from the Borel mass
$M_B$, the threshold value $s_0$, the strange and charm quark
masses, and the various
condensates.
This value is consistent with the experimental mass of the
$X(4500)$~\cite{lhcb}, supporting it to be a $D$-wave
$cs\bar{c}\bar{s}$ tetraquark state of $J^P = 0^+$. It consists of
one $D$-wave ``bad'' diquark and one $S$-wave ``bad'' antidiquark,
having the antisymmetric color structure $[\mathbf{\bar 3_c}]_{cs}
\otimes [\mathbf{3_c}]_{\bar{c}\bar s}$.

\begin{figure*}[hbt]
\begin{center}
\scalebox{0.6}{\includegraphics{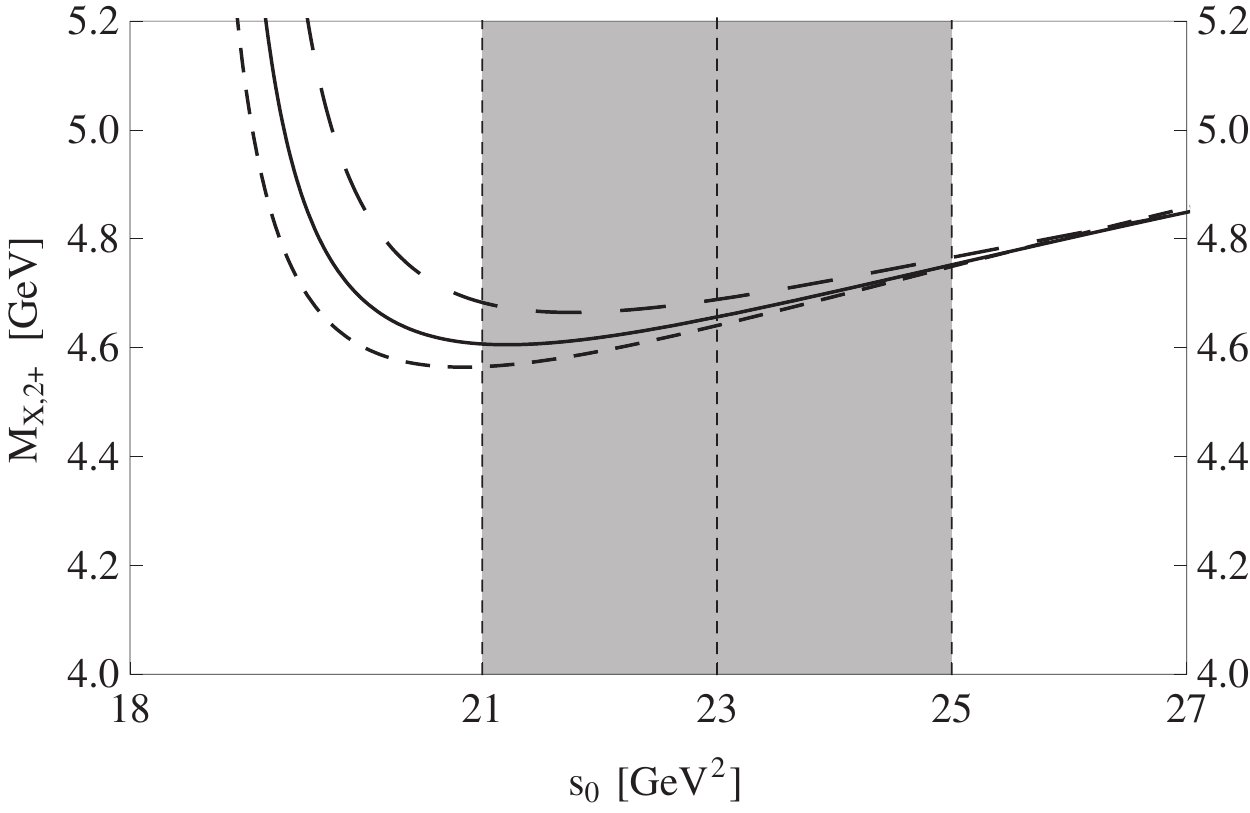}}
\scalebox{0.6}{\includegraphics{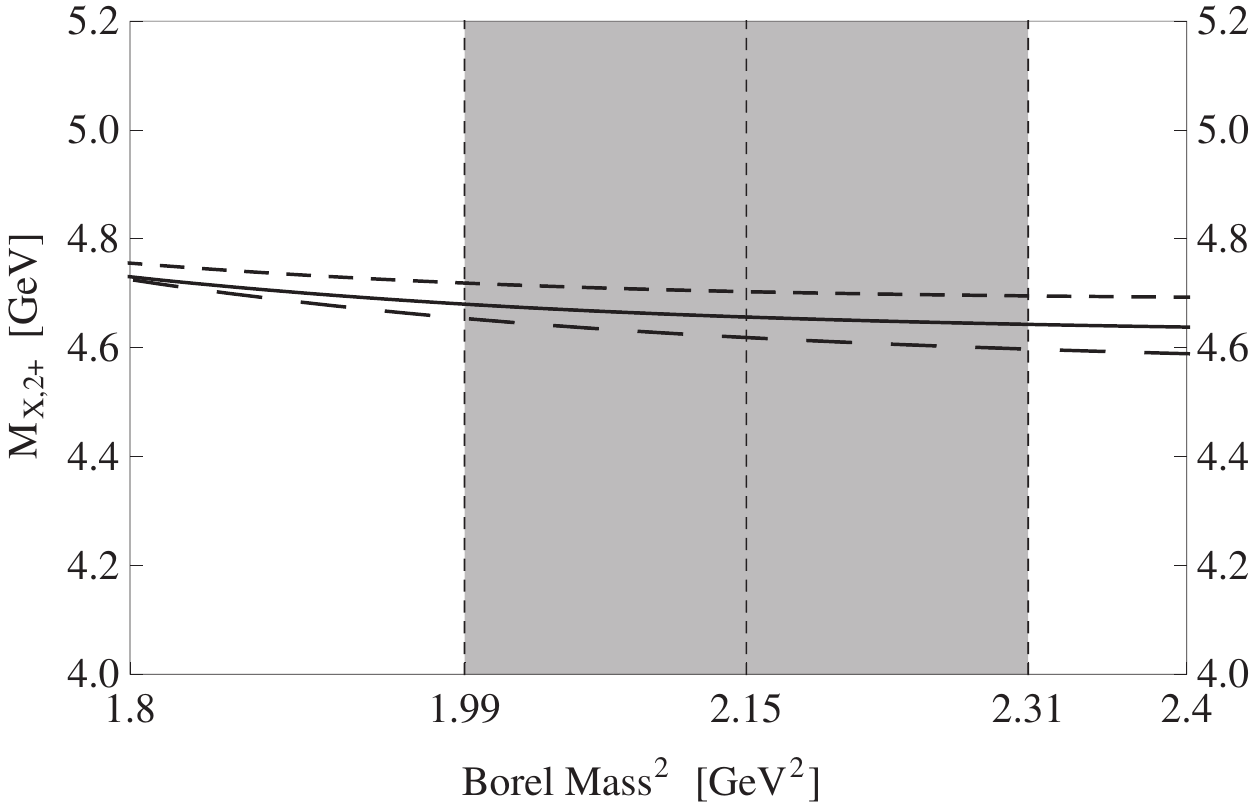}} \caption{The
variation of $M_X$ with respect to the threshold value $s_0$ (left)
and the Borel mass $M_B$ (right), calculated using the current
$J_{2+}$ of $J^P = 0^+$. In the left figure, the long-dashed, solid
and short-dashed curves are obtained by fixing $M_B^2 = 1.95$,
$2.15$ and $2.35$ GeV$^2$, respectively. In the right figure, the
long-dashed, solid and short-dashed curves are obtained for $s_0 =
22$, $23$ and $24$ GeV$^2$, respectively.} \label{fig:j2p}
\end{center}
\end{figure*}

The partner of the $X(4500)$ having the symmetric color structure,
$[\mathbf{6_c}]_{cs} \otimes [\mathbf{ \bar 6_c}]_{\bar{c}\bar s}$,
can be investigated using the current $J_{2+}$. We use this current
to perform sum rule analyses, and show the obtained mass $M_{X,2+}$
in Fig.~\ref{fig:j2p} as a function of $s_0$ and $M_B$. We find that
the $s_0$ dependence is the weakest around $s_0 \sim 21$ GeV$^2$,
and the $M_B$ dependence is the weakest around $s_0 \sim 25$
GeV$^2$. Accordingly, we fix our working regions to be $21$
GeV$^2\leq s_0\leq 25$ GeV$^2$ and $1.99$ GeV$^2$ $\leq M_B^2 \leq
2.31$ GeV$^2$, where the $s_0$ and $M_B$ dependence is both
acceptable. The mass can be extracted to be
\begin{eqnarray}
M_{X,2+} = 4.66^{+0.20}_{-0.14} \mbox{ GeV} \, , \label{mass2+}
\end{eqnarray}
where the central value corresponds to $M_B^2$ = 2.15 GeV$^2$ and
$s_0= 23$ GeV$^2$. We find that there exist Borel windows only when
$s_0 \geq 22$ GeV$^2$, which threshold value is the same as that for
$J_{2-}$. However, if we choose $s_0 = 22$ GeV$^2$, the Borel window
would be quite narrow ($1.99$ GeV$^2$ $\leq M_B^2 \leq 2.03$
GeV$^2$), but the mass extracted would not change much ($M_{X,2+} =
4.66^{+0.36}_{-0.19} \mbox{ GeV}$). The value listed in
Eq.~(\ref{mass2+}) is consistent with the experimental mass of the
$X(4700)$~\cite{lhcb}, suggesting that it can also be interpreted as
a $D$-wave $cs\bar{c}\bar{s}$ tetraquark state of $J^P = 0^+$. It
consists of one $D$-wave diquark and one $S$-wave antidiquark,
having the symmetric color structure $[\mathbf{6_c}]_{cs} \otimes
[\mathbf{ \bar 6_c}]_{\bar{c}\bar s}$.

{\it Conclusion and Discussions.}--- To summarize, we have used the
method of QCD sum rule to investigate the $X(4500)$ and $X(4700)$ of
$J^P = 0^+$ based on the diquark-antidiquark configuration within
the framework of QCD sum rules. We find that the $X(4500)$ and
$X(4700)$ can be both interpreted as $D$-wave tetraquark states with
the quark content $cs\bar{c}\bar{s}$ and $J^P = 0^+$: the $X(4500)$
consists of one $D$-wave ``bad'' diquark and one $S$-wave ``bad''
antidiquark, with the antisymmetric color structure $[\mathbf{\bar
3_c}]_{cs} \otimes [\mathbf{3_c}]_{\bar{c}\bar s}$; the $X(4700)$
consists of similar diquarks, but with the symmetric color structure
$[\mathbf{6_c}]_{cs} \otimes [\mathbf{ \bar 6_c}]_{\bar{c}\bar s}$.
These two interpretations are remarkably similar to those obtained
in Ref.~\cite{Chen:2010ze} that the $X(4140)$ and $X(4274)$ can be
both interpreted as $S$-wave $cs\bar{c}\bar{s}$ tetraquark states of
$J^P = 0^+$, but with opposite color structures.

\begin{figure*}[hbt]
\begin{center}
\scalebox{0.6}{\includegraphics{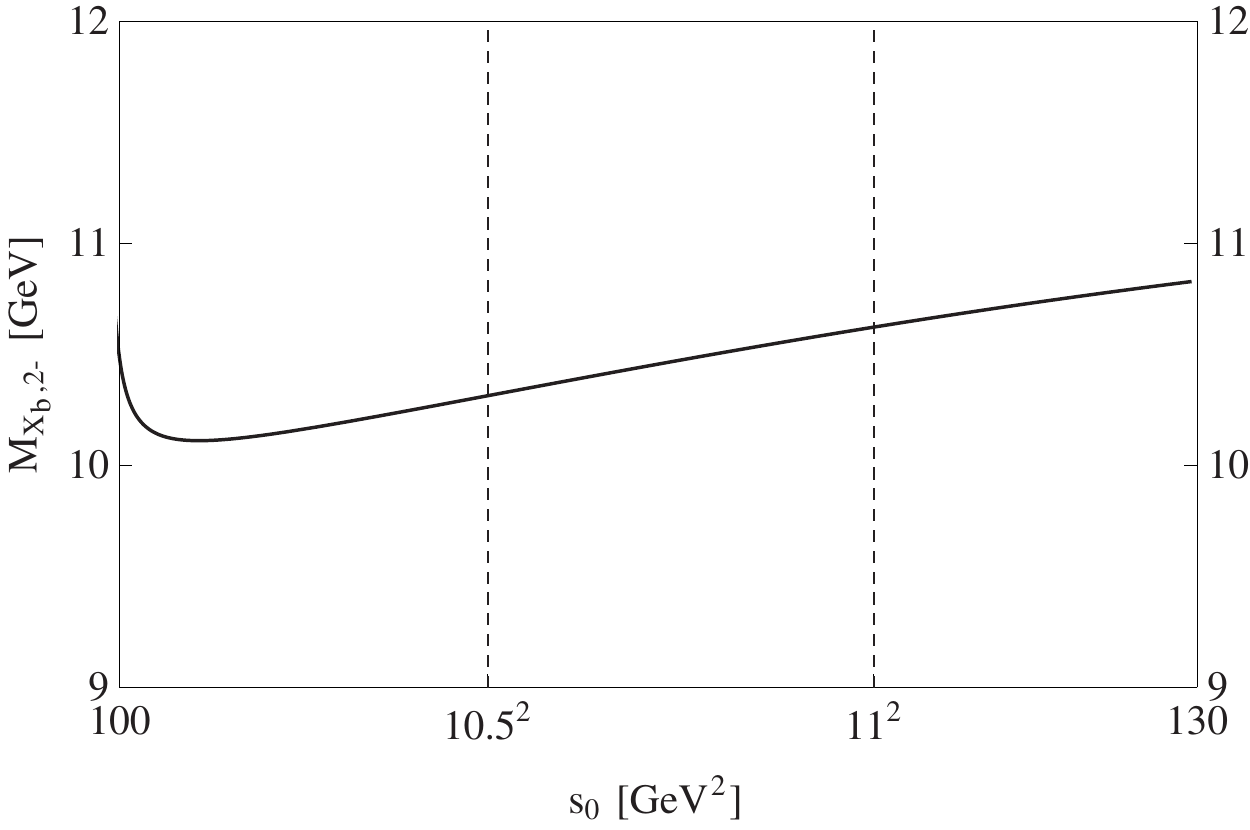}}
\scalebox{0.6}{\includegraphics{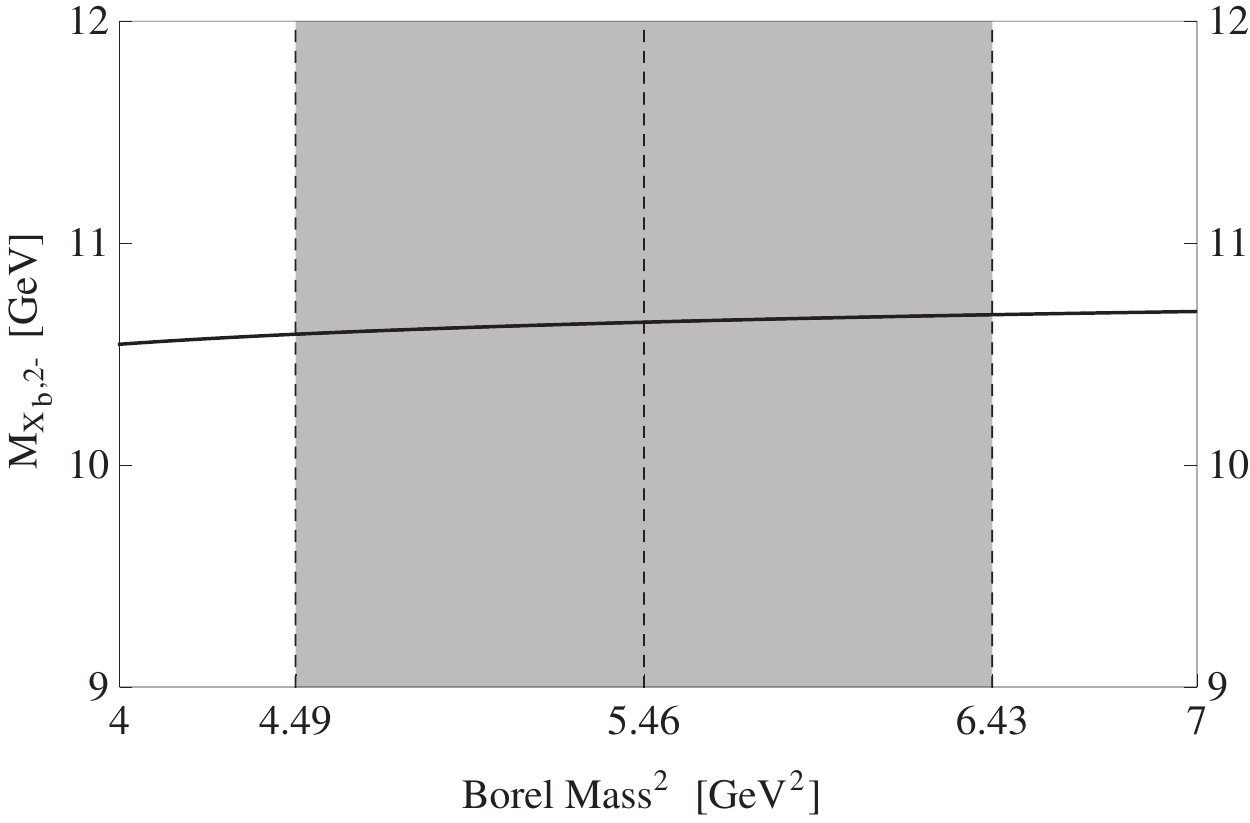}} \caption{The
variation of $M_X$ with respect to the threshold value $s_0$ (left)
and the Borel mass $M_B$ (right), calculated using the current
$J_{2-}(b s \bar b \bar s)$ of $J^P = 0^+$. In the left figure, the
Borel mass $M_B$ is fixed to be $M^2_B= 5.0$ GeV$^2$. In the right
figure, the threshold value $s_0$ is fixed to be $s_0 = 11^2$
GeV$^2$, and the Borel window is $4.49$ GeV$^2$ $\leq M_B^2 \leq
6.43$ GeV$^2$.} \label{fig:jbb}
\end{center}
\end{figure*}

The possible decay channels of the $X(4500)$ and $X(4700)$ can be investigated
by performing the Fierz and color rearrangements on the currents $J_{2\pm}$ and changing
them to mesonic-mesonic structures~\cite{Chen:2016qju,Chen:2006hy,Chen:2007xr}:
\begin{eqnarray}
J_{2\pm} &\rightarrow& \Bigg (
\big [ \bar{s}_a \gamma_{\mu_1} c_b \big ] \big[ \bar{c}_a \gamma_{\mu_2} D_{\mu_3}D_{\mu_4} s_b \pm \{ a \leftrightarrow b \} \big]
\\ \nonumber && \oplus
\big [ \bar{s}_a \gamma_{\mu_1} \gamma_5 c_b \big ] \big[ \bar{c}_a \gamma_{\mu_2} \gamma_5 D_{\mu_3}D_{\mu_4} s_b \pm \{ a \leftrightarrow b \} \big]
\\ \nonumber && \oplus
\big [ \bar{s}_a \sigma_{\mu_1\rho} c_b \big ] \big[ \bar{c}_a \sigma_{\mu_2\rho} D_{\mu_3}D_{\mu_4} s_b \pm \{ a \leftrightarrow b \} \big]
\Bigg )
\\ \nonumber && \times \left( g^{\mu_1 \mu_3} g^{\mu_2 \mu_4} + g^{\mu_1 \mu_4} g^{\mu_2 \mu_3} - g^{\mu_1 \mu_2} g^{\mu_3 \mu_4}/2 \right) \, .
\end{eqnarray}
Besides these structures, their similar/relevant $[\bar s s][\bar c c]$ structures are also possible. Accordingly,
we obtain the possible decay channels of the $X(4500)$ and $X(4700)$ to be
$S$-wave $D_s^{*+} D_s^{*-}$, $D_s^{*+} D_{s1}^{*-}(2860)$, $D_s^{*+} D_{s3}^{*-}(2860)$, $J/\psi \phi$, $J/\psi \phi_3(1850)$,
$P$-wave $D_s^{*+} D_{s0}^{*-}$, $D_s^{*+} D_{s1}^{-}$, $D_s^{*+} D_{s2}^{*-}$,
and $D$-wave $D_s^{*+} D_s^{*-}$ and $J/\psi \phi$, etc.
The $X(4500)$ and $X(4700)$ were observed by LHCb in the $J/\psi \phi$ channel,
which probably contain both $S$-wave and $D$-wave components.
However, the overlap of the $S$-wave $J/\psi \phi$ channel (as well as
the $S$-wave $D_s^{*+} D_s^{*-}$ channel) and the $J_{2\pm}$ (containing the $D$-wave antidiquark) is quite small, which makes the widths of the $X(4500)$ and $X(4700)$
not very large.

To examine these interpretations, we have also studied the bottom
partners of the $X(4500)$ and $X(4700)$ by simply replacing charm
quarks to be bottom quarks. We evaluate their masses using the
bottom quark mass $m_b (m_b) = 4.20 \pm 0.07$ GeV in the
$\overline{\rm MS}$ scheme~\cite{Chen:2010ze,Agashe:2014kda}. We
show the mass obtained using $J_{2-}(bs \bar b\bar s)$ in
Fig.~\ref{fig:jbb} as a function of the threshold values $s_0$ and
the Borel mass $M_B$. There is a mass plateau around $s_0 \sim 10^2$
GeV$^2$, but it depends on the bottom quark mass, which has large
uncertainty~\cite{Agashe:2014kda}. We choose $s_0 = 11^2$
GeV$^2$~\cite{Chen:2010ze} (there exist Borel windows when $s_0 \geq
10.5^2$ GeV$^2$), and the mass obtained is around 10.64 GeV. The
mass obtained using $J_{2+}(bs \bar b\bar s)$ is also around 10.64
GeV. We propose to search for them in the $\Upsilon \phi$ invariant
mass distribution with the running of LHC at 13 TeV and forthcoming
BelleII.

Besides the above dependence on the bottom quark mass, the extracted
masses of the $X(4500)$ and $X(4700)$ in the present work also
depend on the running strange and charm quark masses, which further
depend on the energy scale. Therefore, our results still have some
extra theoretical uncertainties not included in Eqs.~(\ref{mass2-})
and (\ref{mass2+}), and more theoretical and experimental studies
are necessary to understand their internal structures. Especially,
the determination/confirmation of their spin-parity quantum numbers
in experiments can be essential. We also note that the $X(4140)$,
$X(4274)$, $X(4500)$ and $X(4700)$ can have many partner states. If
their interpretations in this letter are correct, their dual partner
states would be quite interesting, such as the $S$-wave scalar and
the $D$-wave axial-vector tetraquark states with the quark content
$cs\bar{c}\bar{s}$. Especially in the diquark-antidiquark
configuration, the $S$-wave scalar $cs\bar{c}\bar{s}$ tetraquark
state consisting of two ``bad'' diquarks is the dual partner state
of both the $X(4140)$ (by replacing one ``good'' diquark by one
``bad'' diquark) and the $X(4500)$ (as its ground state), which may
also exist.

To end this paper, we note that we can also use the $S/P/D$-waves diquarks and antidiquarks to construct many other states, and we plan to use QCD sum rules to systematically study them. Although QCD sum rule studies can not predict their existence, our studies can still be helpful to experimental searching of new exotic hadrons.

\section*{Acknowledgments}
\begin{acknowledgement}
This project is supported by the Natural Sciences and Engineering
Research Council of Canada (NSERC) and the National Natural Science
Foundation of China under Grants 11205011, No. 11375024, No.
11222547, No. 11175073, No. 11575008, and No. 11261130311; the Ministry of Education
of China (the Fundamental Research Funds for the Central
Universities), 973 program. Xiang Liu is also supported by the
National Youth Top-notch Talent Support Program
(``Thousandsof-Talents Scheme").
\end{acknowledgement}



\begin{thebibliography}{99}

\bibitem{Chen:2010ze}
  W.~Chen and S.~L.~Zhu,
  Phys.\ Rev.\ D {\bf 83}, 034010 (2011).

\bibitem{Agashe:2014kda}
  K.~A.~Olive {\it et al.}  [Particle Data Group Collaboration],
  {Review of Particle Physics},
  Chin.\ Phys.\ C {\bf 38}, 090001 (2014).

\bibitem{Jaffe:2004ph}
  R.~L.~Jaffe,
  Phys.\ Rept.\  {\bf 409}, 1 (2005).

\bibitem{Liu:2013waa}
  X.~Liu,
  Chin.\ Sci.\ Bull.\  {\bf 59}, 3815 (2014).

\bibitem{Chen:2016qju}
  H.~X.~Chen, W.~Chen, X.~Liu and S.~L.~Zhu,
  Phys.\ Rept.\  {\bf 639}, 1 (2016).

\bibitem{Aaij:2015tga}
  R.~Aaij {\it et al.} [LHCb Collaboration],
  Phys.\ Rev.\ Lett.\  {\bf 115}, 072001 (2015).

\bibitem{lhcb}
  R.~Aaij {\it et al.} [LHCb Collaboration], arXiv:1606.07895 [hep-ex];
  R.~Aaij {\it et al.} [LHCb Collaboration], arXiv:1606.07898 [hep-ex];
  Talk given by T.~Skwarnicki, on behalf of the LHCb Collaboration at Meson2016, see http://meson.if.uj.edu.pl/indico/event/3/session/1/contribution/ 16/material/slides/0.pdf;
  Talk given by T.~Britton, on behalf of the LHCb Collaboration at APS April Meeting 2016, see https://absuploads.aps.org/presentation.cfm?pid=11733.

\bibitem{Aaltonen:2009tz}
  T.~Aaltonen {\it et al.} [CDF Collaboration],
  Phys.\ Rev.\ Lett.\  {\bf 102}, 242002 (2009).

\bibitem{Aaltonen:2011at}
  T.~Aaltonen {\it et al.} [CDF Collaboration],
  arXiv:1101.6058 [hep-ex].

\bibitem{Liu:2008tn}
  X.~Liu, Z.~G.~Luo, Y.~R.~Liu and S.~L.~Zhu,
  Eur.\ Phys.\ J.\ C {\bf 61}, 411 (2009).

\bibitem{Liu:2009ei}
  X.~Liu and S.~L.~Zhu,
  Phys.\ Rev.\ D {\bf 80}, 017502 (2009)
  Erratum: [Phys.\ Rev.\ D {\bf 85}, 019902 (2012)].

\bibitem{Mahajan:2009pj}
  N.~Mahajan,
  Phys.\ Lett.\ B {\bf 679}, 228 (2009).

\bibitem{Branz:2009yt}
  T.~Branz, T.~Gutsche and V.~E.~Lyubovitskij,
  Phys.\ Rev.\ D {\bf 80}, 054019 (2009).

\bibitem{Ding:2009vd}
  G.~J.~Ding,
  Eur.\ Phys.\ J.\ C {\bf 64}, 297 (2009).

\bibitem{Liu:2010hf}
  X.~Liu, Z.~G.~Luo and S.~L.~Zhu,
  Phys.\ Lett.\ B {\bf 699}, 341 (2011)
  Erratum: [Phys.\ Lett.\ B {\bf 707}, 577 (2012)].

\bibitem{Wang:2011uk}
  Z.~G.~Wang,
  Int.\ J.\ Mod.\ Phys.\ A {\bf 26}, 4929 (2011).

\bibitem{Finazzo:2011he}
  S.~I.~Finazzo, M.~Nielsen and X.~Liu,
  Phys.\ Lett.\ B {\bf 701}, 101 (2011).

\bibitem{He:2011ed}
  J.~He and X.~Liu,
  Eur.\ Phys.\ J.\ C {\bf 72}, 1986 (2012).

\bibitem{HidalgoDuque:2012pq}
  C.~Hidalgo-Duque, J.~Nieves and M.~P.~Valderrama,
  Phys.\ Rev.\ D {\bf 87}, 076006 (2013).

\bibitem{Stancu:2009ka}
  F.~Stancu,
  J.\ Phys.\ G {\bf 37}, 075017 (2010).

\bibitem{Patel:2014vua}
  S.~Patel, M.~Shah and P.~C.~Vinodkumar,
  Eur.\ Phys.\ J.\ A {\bf 50}, 131 (2014).

\bibitem{Molina:2009ct}
  R.~Molina and E.~Oset,
  Phys.\ Rev.\ D {\bf 80}, 114013 (2009).

\bibitem{Branz:2010rj}
  T.~Branz, R.~Molina and E.~Oset,
  Phys.\ Rev.\ D {\bf 83}, 114015 (2011).

\bibitem{Danilkin:2009hr}
  I.~V.~Danilkin and Y.~A.~Simonov,
  Phys.\ Rev.\ D {\bf 81}, 074027 (2010).

\bibitem{vanBeveren:2009dc}
  E.~van Beveren and G.~Rupp,
  arXiv:0906.2278 [hep-ph].

\bibitem{Albuquerque:2009ak}
  R.~M.~Albuquerque, M.~E.~Bracco and M.~Nielsen,
  Phys.\ Lett.\ B {\bf 678}, 186 (2009).

\bibitem{Zhang:2009st}
  J.~R.~Zhang and M.~Q.~Huang,
  J.\ Phys.\ G {\bf 37}, 025005 (2010).

\bibitem{Wang:2009ue}
  Z.~G.~Wang,
  Eur.\ Phys.\ J.\ C {\bf 63}, 115 (2009).

\bibitem{Wang:2009ry}
  Z.~G.~Wang, Z.~C.~Liu and X.~H.~Zhang,
  Eur.\ Phys.\ J.\ C {\bf 64}, 373 (2009).

\bibitem{Chen:2006zh}
  H.~X.~Chen, A.~Hosaka and S.~L.~Zhu,
  Phys.\ Lett.\ B {\bf 650}, 369 (2007).

\bibitem{Chen:2015moa}
  H.~X.~Chen, W.~Chen, X.~Liu, T.~G.~Steele and S.~L.~Zhu,
  Phys.\ Rev.\ Lett.\  {\bf 115}, 172001 (2015).

\bibitem{Chen:2016ymy}
  H.~X.~Chen, D.~Zhou, W.~Chen, X.~Liu and S.~L.~Zhu,
  Eur.\ Phys.\ J.\ C {\bf 76}, 602 (2016).

\bibitem{Chen:2006hy}
  H.~X.~Chen, A.~Hosaka and S.~L.~Zhu,
  Phys.\ Rev.\ D {\bf 74}, 054001 (2006).

\bibitem{Chen:2007xr}
  H.~X.~Chen, A.~Hosaka and S.~L.~Zhu,
  Phys.\ Rev.\ D {\bf 76}, 094025 (2007).

\bibitem{Maiani:2004vq}
  L.~Maiani, F.~Piccinini, A.~D.~Polosa and V.~Riquer,
  Phys.\ Rev.\ D {\bf 71}, 014028 (2005).

\bibitem{Kleiv:2013dta}
  R.~T.~Kleiv, T.~G.~Steele, A.~Zhang and I.~Blokland,
  Phys.\ Rev.\ D {\bf 87}, 125018 (2013).

\bibitem{Lebed:2016yvr}
  R.~F.~Lebed and A.~D.~Polosa,
  Phys.\ Rev.\ D {\bf 93}, 094024 (2016).

\bibitem{Shifman:1978bx}
  M.~A.~Shifman, A.~I.~Vainshtein and V.~I.~Zakharov,
  Nucl.\ Phys.\ B {\bf 147}, 385 (1979).

\bibitem{Reinders:1984sr}
  L.~J.~Reinders, H.~Rubinstein and S.~Yazaki,
  Phys.\ Rept.\  {\bf 127}, 1 (1985).

\bibitem{Nielsen:2009uh}
  M.~Nielsen, F.~S.~Navarra and S.~H.~Lee,
  Phys.\ Rept.\  {\bf 497}, 41 (2010).

\bibitem{colangelo}
  P.~Colangelo and A.~Khodjamirian, {\it ``At the Frontier of
  Particle Physics/Handbook of QCD''} (World Scientific,
  Singapore, 2001), Volume 3, 1495.

\bibitem{Narison:2002pw}
  S.~Narison,
  Camb.\ Monogr.\ Part.\ Phys.\ Nucl.\ Phys.\ Cosmol.\  {\bf 17}, 1 (2002).

\bibitem{Zhou:2014ytp}
  D.~Zhou, E.~L.~Cui, H.~X.~Chen, L.~S.~Geng, X.~Liu and S.~L.~Zhu,
  Phys.\ Rev.\ D {\bf 90}, 114035 (2014).

\bibitem{Zhou:2015ywa}
  D.~Zhou, H.~X.~Chen, L.~S.~Geng, X.~Liu and S.~L.~Zhu,
  Phys.\ Rev.\ D {\bf 92}, 114015 (2015).

\bibitem{Chen:2011qu}
  W.~Chen, Z.~X.~Cai and S.~L.~Zhu,
  Nucl.\ Phys.\ B {\bf 887}, 201 (2014).

\bibitem{Eidemuller:2000rc}
  M.~Eidemuller and M.~Jamin,
  Phys.\ Lett.\ B {\bf 498}, 203 (2001).

\bibitem{Gimenez:2005nt}
  V.~Gimenez, V.~Lubicz, F.~Mescia, V.~Porretti and J.~Reyes,
  Eur.\ Phys.\ J.\ C {\bf 41}, 535 (2005).

\bibitem{Jamin:2002ev}
  M.~Jamin,
  Phys.\ Lett.\ B {\bf 538}, 71 (2002).

\bibitem{Ioffe:2002be}
  B.~L.~Ioffe and K.~N.~Zyablyuk,
  Eur.\ Phys.\ J.\ C {\bf 27}, 229 (2003).

\bibitem{Ovchinnikov:1988gk}
  A.~A.~Ovchinnikov and A.~A.~Pivovarov,
  Sov.\ J.\ Nucl.\ Phys.\  {\bf 48}, 721 (1988)
  [Yad.\ Fiz.\  {\bf 48}, 1135 (1988)].

\bibitem{Narison:1996fm}
  S.~Narison,
  Nucl.\ Phys.\ B {\bf 509}, 312 (1998).

\bibitem{Chen:2014vha}
  H.~X.~Chen, E.~L.~Cui, W.~Chen, T.~G.~Steele and S.~L.~Zhu,
  Phys.\ Rev.\ C {\bf 91}, 025204 (2015).

\end{thebibliography}
\end{document}